# Explainable Artificial Intelligence for Economic Time Series: A Comprehensive Review and a Systematic Taxonomy of Methods and Concepts

Agustín García-García [1], Pablo Hidalgo [2], Julio E. Sandubete [2,*]

[1] Faculty of Economics and Business Studies, Universidad de Extremadura, 06006 Badajoz, Spain
[2] Faculty of Law, Business and Government, Universidad Francisco de Vitoria, 28223 Madrid, Spain.
* Correspondence: je.sandubete@ufv.es

**Abstract:** Explainable Artificial Intelligence (XAI) is increasingly required in computational economics, where machine-learning forecasters can outperform classical econometric models but remain difficult to audit and use for policy. This survey reviews and organizes the growing literature on XAI for economic time series, where autocorrelation, non-stationarity, seasonality, mixed frequencies, and regime shifts can make standard explanation techniques unreliable or economically implausible. We propose a taxonomy that classifies methods by *(i)* explanation mechanism: propagation-based approaches (e.g., Integrated Gradients, Layer-wise Relevance Propagation), perturbation and game-theoretic attribution (e.g., permutation importance, LIME, SHAP), and function-based global tools (e.g., Accumulated Local Effects); *(ii)* time-series compatibility, including preservation of temporal dependence, stability over time, and respect for data-generating constraints. We synthesize time-series-specific adaptations such as vector- and window-based formulations (e.g., Vector SHAP, WindowSHAP) that reduce lag fragmentation and computational cost while improving interpretability. We also connect explainability to causal inference and policy analysis through interventional attributions (Causal Shapley values) and constrained counterfactual reasoning. Finally, we discuss intrinsically interpretable architectures (notably attention-based transformers) and provide guidance for decision-grade applications such as nowcasting, stress testing, and regime monitoring, emphasizing attribution uncertainty and explanation dynamics as indicators of structural change.

**Keywords:** Explainable AI; economic time series; feature attribution; SHAP; causal explainability.

## 1. Introduction: The Imperative of Interpretability in Computational Economics

The integration of Machine Learning (ML) and Deep Learning (DL) methods into economic and financial modeling has triggered a fundamental paradigm shift (Mullainathan & Spiess, 2017; Varian, 2014). Historically, econometrics has relied on linear models, such as multiple linear regression or Vector Autoregressions (VAR) (Lütkepohl, 2005), where transparency is inherent: a coefficient β provides a direct and global view of marginal effects (a unit increase in interest rates results in a decrease of β in investment, ceteris paribus) (Wooldridge, 2010). However, the rigidity of these models often fails to capture the complex nonlinear dynamics and higher-order interactions characteristic of modern financial markets and global supply chains (Koop et al., 1996).

In contrast, "black-box" models such as Recurrent Neural Networks (RNNs) or Transformers have demonstrated superior capacity to approximate these complex functions (Makridakis et al., 2018; Zerveas et al., 2021), but at the cost of obscuring the decision boundary (Rudin, 2019). This trade-off between accuracy and transparency has become unsustainable in the face of strict regulatory frameworks, such as the "Right to Explanation" under the General Data Protection Regulation (GDPR) in Europe (Goodman & Flaxman, 2017) or Basel III/IV risk management guidelines, which require financial institutions to justify capital requirements and credit decisions. Consequently, Explainable Artificial Intelligence (XAI) has emerged not as a luxury, but as a strategic and legal necessity (Barredo Arrieta et al., 2020), aiming to bridge the cognitive gap between the predictive power of algorithms and the human need for causal understanding (Doshi-Velez & Kim, 2017).

The challenge of applying XAI to economics lies in the idiosyncratic nature of economic time series. Unlike static image or tabular data, economic series exhibit autocorrelation, stochastic trends, seasonality, and structural regime shifts (Hamilton, 1994; Stock, 2020). Standard explainability methods that assume feature independence can disrupt the temporal structure of the data (Molnar, 2025), creating unrealistic "counterfactual" scenarios that violate the arrow of time or the coherence of the economic state (Tonekaboni et al., 2019). Recent literature has therefore begun to adapt general XAI techniques to the specific constraints of economic inference (Belle & Papantonis, 2021).

## 2. Propagation and Factor Attribution Mechanisms in Financial Models

Within the broad spectrum of explainability techniques, propagation-based methods represent an "intrinsic" or white-box approach, leveraging the internal model architecture to trace the flow of information.

### 2.1 Fundamentals of Relevance Propagation

Propagation techniques, mainly used in Artificial Neural Networks (ANNs), exploit the differentiability and multiplicative composition of network layers to compute feature importance by tracking contributions from the model output back to its input. This approach is based on the theoretical premise that the final prediction can be decomposed into the sum of the relevance of the input neurons.

A foundational method in this category is Integrated Gradients (Sundararajan et al., 2017), which accumulates gradients along a linear path from a baseline to the current input. This method satisfies the completeness axiom, ensuring that the sum of attributions equals exactly the difference between the model prediction and the baseline prediction a crucial property for financial auditing, where each basis point change in a risk prediction must be justified.

Among the most representative and sophisticated methods is Layer-Wise Relevance Propagation (LRP) (Bach et al., 2015). Unlike raw gradients, which can be noisy or suffer from saturation (where changes in input no longer affect output due to the nature of activation functions), LRP propagates relevance scores layer by layer using conservation rules, redistributing evidence of the prediction backward.

### 2.2 Application in Deep Factor Networks

In finance, applying LRP has enabled significant advances in understanding asset valuation models. Traditional factor models (such as the Fama-French three-factor model) assume static linear relationships between asset returns and risk factors like size, value, or momentum. However, Nakagawa et al. (2019) challenged this orthodoxy by applying LRP to a Deep Factor Network.

Their study used LRP to evaluate the dynamic contribution of multiple factors (risk, value, size, among others) in a neural network model trained to predict stock returns. Results revealed substantial divergences compared to traditional linear sensitivity measures, such as Kendall or Spearman correlations. While linear correlations suggested a constant exposure to certain factors, LRP analysis showed that the neural network captured nonlinear, regime-dependent exposures. For instance, the "Value" factor could have a dominant positive influence during economic recovery periods but become irrelevant or negative during liquidity crises. This ability of LRP to visualize factor importance locally and temporally allowed the model to achieve higher predictive accuracy than linear models while maintaining interpretability aligned with financial intuition but enriched with nonlinear nuances (Nakagawa et al., 2019).

## 3. Perturbation Techniques and Temporal Sensitivity Analysis

While propagation-based methods require access to the model's internal gradients, perturbation techniques operate under a model-agnostic approach, treating the predictive function f(x) as a black box.

**3.1 Classical Perturbation Methodologies**

Perturbation techniques estimate variable importance by altering a single feature and comparing the modified prediction with the original. This approach has roots in classical statistics and machine learning, including foundational methods proposed by Breiman (2001), such as permutation importance in Random Forests. In this framework, values of a feature are randomly shuffled across the dataset to break their association with the target variable while maintaining their marginal distribution; the resulting drop in model accuracy quantifies the variable's importance.

The spectrum of perturbation techniques includes classical methods such as adding noise (introducing stochastic variations) and permutation importance. In financial time series, these techniques have been extensively applied to sequential architectures like RNNs and LSTMs.

**3.2 Complementarity in Financial Series Analysis**

A comparative study by Cascarino et al. (2022) on financial time series revealed that perturbation and propagation methods are not mutually exclusive but complementary. Their analysis on RNNs showed that methods like integrated gradients are superior for capturing temporal importance (identifying when a critical event occurred in the observation window, e.g., a volatility spike three days ago), while permutation is more effective for assessing feature relevance (identifying which variable, e.g., Price vs. Volume, drove the decision).

This distinction is vital in market microstructure. Freeborough (2022) applied advanced perturbation techniques, including SHAP, to predict cryptoassets and traditional markets. Studies consistently conclude that price-related variables exert significantly higher predictive influence than volume variables. Moreover, temporal decay in importance was observed: recent values (t-1, t-2) carry much greater predictive weight than distant lags, aligning with the weak form of the Efficient Market Hypothesis, which suggests that past information is quickly incorporated into prices (Freeborough, 2022). However, simple permutation in time series can be problematic because it destroys autocorrelation structure, leading to the development of more sophisticated methods.

**3.3 SHAP (SHapley Additive exPlanations): The Gold Standard in Economics**

SHAP has become an advanced explainability method based on perturbation and cooperative game theory. It assigns each feature an importance value based on its marginal contribution to the model output, averaged over all possible feature coalitions.

**3.3.1 Theoretical Foundations and Economic Properties**

Developed by Lundberg et al. (2020), SHAP evaluates the effect of each variable by considering all combinations in which it may be present or absent. SHAP is particularly attractive for economics and risk auditing because it satisfies three axiomatic properties from game theory:

- Local Accuracy (Efficiency): The sum of SHAP values of all features exactly equals the difference between the model prediction for a specific instance and the baseline prediction. Economically, this ensures that 100% of a predicted deviation (e.g., "Why is the forecasted inflation 2% above the historical mean?") is attributed to specific drivers without unexplained residuals.
- Consistency: If a model change such that a feature's marginal contribution increases or stays the same, its SHAP value will not decrease. This enables reliable comparisons between model versions (e.g., Q1 Model vs. Q2 Model).

- Missingness: A feature that is missing or has no impact on the model receives a SHAP value of zero.

Variants adapted to different model architectures include Kernel SHAP (a model-agnostic approximation based on locally weighted linear regression) and TreeSHAP, optimized for tree-based models like XGBoost and Random Forest, which are ubiquitous in credit scoring.

### 3.3.2 Global and Local Interpretation

SHAP's main advantage lies in its duality: it provides both local feature importance (explaining an individual prediction) and global feature importance (aggregating absolute SHAP values across the dataset). This allows transparent interpretation of the positive or negative contribution of each variable to predictions (Weng et al., 2022).

For example, a global analysis might reveal that "Interest Rates" are the most important factor for predicting mortgage defaults overall. However, a local SHAP analysis could show that, for a specific client, the key factor was an unusually high "Debt-to-Income Ratio." This granularity is essential for personalized portfolio management and lending fairness.

### 3.4 LIME and Temporal Stability Challenges

LIME (Local Interpretable Model-agnostic Explanations), proposed by Ribeiro et al. (2016), explains predictions of any black-box model by locally approximating its behavior with an interpretable model, usually sparse linear regression (Lasso).

### 3.4.1 Mechanism and Limitations

LIME perturbs input data around the instance of interest (generating new samples with noise), obtains predictions from the complex model for these samples, and fits a surrogate model weighted by proximity. This allows identification of how each variable influences the prediction in that specific neighborhood.

However, its main limitation is that it only provides local explanations, without the ability to generalize globally with the mathematical consistency offered by SHAP (Ghosh et al., 2023). Additionally, defining the appropriate local neighborhood is particularly challenging in time series. What does "local" mean in time? Are temporally adjacent points considered, or points with similar past values? Standard LIME perturbations (e.g., adding Gaussian noise) can create synthetic data points that violate temporal dependencies and autocorrelation structure (e.g., creating an inverted yield curve that is mathematically impossible given prior inputs).

This instability where the explanation for time t and t+1 may vary drastically even if the underlying model is stable has motivated the development of specific variants. A notable example is TS-MULE (Schlegel et al., 2021), which adapts LIME to segment time series into meaningful windows or frequency components rather than perturbing individual points, preserving the temporal structure more effectively.

### 3.5 Accumulated Local Effects (ALE) plots

Accumulated Local Effects (ALE) plots are a model-agnostic, post-hoc explainability technique designed to quantify how individual features influence a fitted predictive function while accounting for the empirical distribution and dependencies among inputs — a common characteristic of economic time series data. Unlike Partial Dependence Plots (PDPs), which average predictions over the marginal distribution of other features and risk extrapolating into unrealistic regions of the feature space, ALE computes local changes in the model output within small intervals of a feature's domain and then accumulates these changes to produce an interpretable summary curve. This approach mitigates bias from correlated predictors and yields a more faithful estimate of feature effects in high-dimensional, autocorrelated contexts such as macroeconomic or financial time series (Apley & Zhu, 2020; Friedman, 2001; Molnar, 2025).

In economic and financial forecasting research, ALE plots have been used both to illuminate the functional relationships learned by complex machine learning models and to complement traditional econometric reasoning. For example, in volatility and realized variance forecasting, ALE visualizations help identify how lagged measures and macroeconomic predictors affect forecasts, revealing patterns such as thresholds, regime shifts, or nonlinear sensitivities that would remain hidden under linear models (Christensen et al., 2023). More broadly, when machine-learning methods such as gradient boosting machines, random forests, or neural networks are compared with classical benchmarks (e.g., AR, HAR, or HAR-X models), ALE provides interpretable insights into how influential predictors contribute to model outputs, beyond mere variable importance rankings (Goldstein et al., 2015; Molnar, 2025).

This emphasis on interpretability within economic time series aligns with broader trends in Explainable Artificial Intelligence (XAI). Recent work in multi-task similarity and explainability explicitly integrates ALE-like techniques to quantify and visualize feature contributions across tasks or predictive horizons; for instance, an explainable multi-task similarity measure proposed in the context of XAI leverages local, accumulated effects to ensure transparent interpretation of model behavior across correlated tasks and time series outputs, underscoring the practical value of ALE-based explanations in complex predictive (Hidalgo & Rodriguez, 2025). Such applications illustrate how ALE extends beyond univariate effect plots to support richer diagnostic insights in both supervised forecasting and similarity analysis when dealing with temporally structured economic data.

Within the taxonomy of XAI methods applied to economic time series, ALE plots occupy an important position among global, model-agnostic, function-based explanation tools. While ALE provides associational rather than causal interpretations, its robustness to correlated inputs and scalability to high-complexity models make it a practical and theoretically grounded component of the explainability toolkit for economic forecasting. By enabling transparent model assessment, hypothesis generation, and communication of machine-learning results to domain experts, ALE plots help bridge the gap between predictive performance and interpretability a key challenge in economic and financial applications of AI (Apley & Zhu, 2020; Molnar, 2025).

**4. Vector and Temporal Adaptations: Overcoming the Curse of Dimensionality**

The standard application of SHAP and LIME to economic time series often ignores the sequential dependence of observations. In econometric and ML models (such as VAR or LSTM), the input vector often consists of multiple lags of various variables (e.g., GDP at t-1, t-2, …, t-12). Treating each lag as an independent "player" in the Shapley game leads to two critical problems: exponential computational explosion and fragmentation of the economic explanation.

**4.1 Vector SHAP: Efficiency in Lag Structures**

It is often more economically meaningful to know the total contribution of "Interest Rates" (the vector of all its lags) than the individual contribution of "Interest Rate at t-5." Recent research has proposed Vector SHAP, a methodology that redefines the "players" in the Shapley game as lag vectors for each variable rather than individual scalar lags.

Instead of permuting $x_{\{t-1\}}$, $x_{\{t-2\}}$ independently, Vector SHAP permutes the entire historical vector $x_{t-k:t}$ of a macroeconomic variable. This drastically reduces problem dimensionality, allowing much faster computation. Theoretically, Vector SHAP satisfies vectorial local accuracy and vectorial consistency. Empirically, in realized volatility prediction for stock indices like KOSPI, Vector SHAP provided consistent importance rankings across global regions (North America, Europe, Asia) with superior computational efficiency compared to standard SHAP, while also mitigating sampling sensitivity.

**4.2 WindowSHAP: Detecting Temporal Regimes**

WindowSHAP addresses dependency issues by focusing on the temporal dimension. It partitions the sequence into time windows (e.g., consecutive days or weeks) and treats each window as a unified feature.

Advanced versions, such as Dynamic WindowSHAP, adaptively size the windows based on feature importance, grouping uninformative time steps to save computational resources while maintaining high resolution for critical events. This is particularly valuable for detecting structural changes or identifying "crisis windows." For example, when predicting sovereign default, WindowSHAP can isolate whether the model is reacting to gradual deterioration over the past year (long-window importance) or a sudden shock in the last week (short-window importance), providing analysts with a clear temporal narrative.

**5. Causality and Inference in Economic XAI**

A critical limitation of standard post-hoc XAI (including SHAP) is that it describes correlations used by the model, not necessarily causal relationships. In economics, distinguishing a variable that predicts an outcome from one that causes it is fundamental for policy design. A model may correctly predict that "more police" correlates with "more crime" (due to resource allocation to problem areas) but inferring that reducing police will reduce crime would be a disastrous causal error.

**5.1 Causal Shapley Values**

Standard SHAP values condition on observed features (conditional expectation E). If features are correlated (e.g., Inflation and Interest Rates), the SHAP value for Interest Rates will absorb part of Inflation's effect, leading to ambiguous attributions. Causal Shapley Values integrate Judea Pearl's Do-Calculus into the Shapley framework. Instead of conditional expectations, they use interventional expectations: $v(S) = E$.

This formulation effectively "cuts" the causal link from parents to intervened variables in the underlying causal graph, ensuring that credit assigned to a feature reflects its causal influence, not spurious correlation.

A major innovation of Causal SHAP is the ability to decompose the total effect into direct effects (the immediate impact of the variable on the target) and indirect effects (mediated through other variables). For example, in a model predicting "Consumer Spending," a tax cut (Feature A) increases Disposable Income (Feature B), which in turn increases Spending. Standard SHAP might attribute most of the effect to Disposable Income if it is the closest predictor. Causal SHAP attributes "root cause" credit to the Tax Cut, recognizing that the income increase was a downstream effect. This distinction is vital for fiscal policy simulation.

**5.2 Challenges in Causal Inference with Observational Data**

Implementing Causal SHAP requires knowledge (or discovery) of the causal graph (Directed Acyclic Graph, DAG). In econometrics, this aligns with structural identification. When the causal graph is unknown, causal discovery algorithms or expert knowledge (priors) are necessary. Recent methods such as Asymmetric Shapley Values (ASV) relax the symmetry axiom to incorporate partial causal information, allowing more nuanced explanations that respect the temporal direction of economic variables.

**6. Counterfactual Analysis and Policy Simulation**

Counterfactuals "What would have happened to Y if X had been different?" are the basis of economic reasoning. In ML, counterfactual explanations seek the minimal change in inputs required to alter a prediction (e.g., "If your income were $5,000 higher, your loan would be approved").

**6.1 Comparison with Impulse-Response Functions (IRFs)**

In traditional Structural VAR (SVAR) models, Impulse-Response Functions (IRFs) serve as the primary counterfactual tool, tracing variable trajectories after an exogenous shock. ML-based counterfactuals offer a more flexible and nonlinear alternative but lack SVAR's structural guarantees. A risk with pure ML counterfactuals is generating "adversarial examples" that are mathematically valid to fool the model but economically absurd (e.g., sustained high inflation with zero interest rates and zero unemployment in a non-Keynesian model). Recent research emphasizes using "impulse-constrained" or manifold-constrained counterfactuals, forcing the ML model to respect historical correlation structure or specific structural constraints when generating hypothetical scenarios. This bridges the flexibility of ML with economic theory rigor, allowing central banks to simulate stress-testing scenarios consistent with observed economic history.

**6.2 Application in Policy Evaluation**

Counterfactuals are essential for assessing the impact of regulatory interventions. For instance, when evaluating a new capital regulation, a counterfactual XAI model can estimate the default probability of a bank under the "no regulation" scenario versus the real scenario. The difference represents the causal treatment effect, aligning with Rubin's potential outcomes framework widely used in microeconometrics.

**7. Detection of Structural Changes and Economic Regimes**

Economic time series are rarely stationary; they are subject to regime changes (e.g., expansion vs. recession, pre- vs. post-COVID). An XAI method providing a single global importance score for "Volatility" can be misleading if volatility is predictive only during crises but is noise during stability.

**7.1 Monitoring Structural Breaks with XAI**

Explainability scores themselves can be treated as time series to detect structural breaks. By monitoring SHAP values over time, analysts can identify points where the model's reasoning structure changes. If the SHAP value for "Oil Prices" suddenly rises from near zero to a dominant driver, this signals a regime shift (e.g., the 1973 Oil Shock), even if the model's predictive accuracy remains stable.

This aligns with classical econometric tests like Chow or CUSUM but offers greater granularity by attributing the break to specific variables. In anomaly detection models based on Autoencoders, a peak in reconstruction error indicates that the current economic state follows "different rules" than the training period. Decomposing this error using SHAP reveals which variables are deviating from historical norms (e.g., "Unemployment" behaving anomalously during COVID-19 lockdown).

**7.2 Anchors and Regime-Based Rules**

Anchors (Ribeiro et al., 2018) is a local rule-based explanation method that finds "sufficient conditions" for a prediction (e.g., "IF Inflation > 5% AND Unemployment < 4%, THEN Probability of Rate Hike is High"). In the context of Temporal Fusion Transformers (TFT) for US inflation forecasting, recent studies have used Anchors to extract interpretable IF-THEN rules from deep learning models. These rules translate opaque neural network weights into critical policy-relevant events. For example, identifying that a specific confluence of "Supply Chain Pressure Index >

1.5" and "Commodity Prices > X" creates a "sufficient condition" for an inflation spike allows central bankers to monitor these thresholds explicitly. This approach is particularly powerful with mixed-frequency data (combining daily financial data with monthly macroeconomic data), precisely signaling which high-frequency shock triggers a low-frequency forecast change.

## 8. Intrinsic Interpretability: Attention Mechanisms and Transformers

While model-agnostic methods (SHAP, LIME) are versatile, they can be computationally inefficient and external to the decision process. Modern "gray-box" architectures, particularly Transformers, offer intrinsic interpretability through Attention Mechanisms.

### 8.1 Temporal Fusion Transformers (TFT) in Macroeconomics

The TFT architecture is explicitly designed for multi-horizon forecasting with interpretability. It uses: *(i)* Variable Selection Networks. Components that weight input variables at each time step, providing built-in feature importance measures; *(ii)* Temporal Self-Attention. Attention weights ($\alpha t,\tau$ indicate how much the forecast at time t+h depends on historical observation at time τ.

Visualizing these attention maps allows economists to see the model's "memory." Does the model look back 12 months (seasonal dependence) or only 1 month (momentum)? In retail demand forecasting, TFTs have been shown to dynamically shift attention from "Past Sales" to "Calendar Events" (e.g., holidays) exactly when expected, validating the model's economic logic.

### 8.2 Natural Language Processing (NLP) in Central Banks

Central banks increasingly use NLP to analyze communication and market sentiment. The "black box" here are large language models (LLMs) classifying a speech as "Hawkish" or "Dovish."

XAI methods visualize attention heads to see which words drive classification. For example, in a Federal Reserve statement, the model may focus heavily on words like "transitory" or "watchful." The Deutsche Bundesbank uses an AI system called MILA (Monetary-Intelligent Language Agent), where sentence-level analysis is aggregated and XAI ensures transparency by highlighting the specific phrases that led to a "negative sentiment" score. This allows human experts to validate the machine reading of monetary policy nuances, creating an essential feedback loop for oversight.

## 9. XAI in Macroeconomic Nowcasting Workflows

Nowcasting (predicting current or near-term GDP) requires handling "ragged edge" data (asynchronous releases). A "decision-grade" nowcasting workflow integrates XAI not just out of curiosity, but for reliability audits. The recent literature proposes a framework including:

- Vintage Management: Explanations must be relative to the information set available in that specific vintage. A SHAP score for "Industrial Production" may be high in the final vintage but zero in an early vintage (before data release).
- Uncertainty Quantification: Point estimates of feature importance are insufficient. Using block bootstrapping, analysts can generate confidence intervals for SHAP values. If the SHAP value for "Retail Sales" has a wide interval crossing zero, the model is uncertain about its contribution. This signals policymakers that the nowcast is fragile and potentially driven by noise rather than a clear economic signal.
- Sign Coherence: Economic theory imposes sign constraints (e.g., higher interest rates should essentially dampen demand). XAI is used to flag sign violations. If a neural network assigns a positive SHAP value to "Interest

Rates" for predicting an increase in "Inflation," investigation is warranted: is it capturing a long-term "Fisher Effect," or is it a spurious correlation in the training sample?

## 10. Conclusion

The integration of Explainable AI into economic time series analysis represents a maturation of the field, moving from raw predictive power to reliable, causal, interpretable intelligence. While foundational propagation and perturbation methods (like LRP and SHAP) provide the basis for feature attribution, domain-specific adaptations such as Vector SHAP, WindowSHAP, and Causal Shapley Values address the technical challenges of temporal dependence and inherent causality in economics.

Modern architectures like Transformers and causal inference techniques are making complex "black boxes" transparent by design or through rigorous post-hoc scrutiny. For economists, these tools offer a new lens: not only to forecast the future but to diagnose the structural dynamics of the present. Whether decomposing a crisis signal in a nowcasting model, auditing bias in credit allocation, or analyzing monetary policy semantics, XAI has become an indispensable component of the modern econometric toolkit. The future lies in human-in-the-loop systems, where XAI not only explains AI to humans but allows experts to impose economic theory constraints back onto AI, creating a symbiotic relationship between data-driven flexibility and theoretical rigor.

**Table 1: Comparison of XAI Methods for Economic Time Series**

| Method | Type | Main Mechanism | Advantage in Economics | Limitation |
|---|---|---|---|---|
| LRP | Propagation | Layer-wise relevance decomposition (Gradients) | Detects nonlinear factor exposures (Deep Factors) | Requires access to internal architecture (White box) |
| Integrated Gradients | Propagation | Gradient accumulation from baseline | Satisfies completeness axiom; good for auditing | Sensitive to baseline choice |
| SHAP | Perturbation | Shapley values (Game Theory) | Strong theoretical properties (Consistency, Additivity) | Computationally expensive; assumes feature independence |
| Vector SHAP | Perturbation | Permutation of lag vectors | Efficient for models with many lags (VAR/LSTM) | Lower temporal granularity within the vector |
| Causal SHAP | Causal | Do-calculus (Intervention) | Distinguishes cause from correlation; separates direct/indirect effects | Requires known or discovered causal DAG |
| LIME | Perturbation | Local surrogate linear model | Intuitive and simple to implement | Unstable in time series; lacks global consistency |
| ALE plots | Perturbation | Accumulated local prediction differences | Handles correlated economic predictors | Associational, sensitive to binning |
| Anchors | Rules | IF-THEN sufficient condition rules | Generates clear policy diagnostics (inflation thresholds) | May be too conservative in rule coverage |
| TFT Attention | Intrinsic | Temporal attention weights | Visualizes model "memory" and seasonality changes | Specific to Transformer architecture |